\documentclass[aps,pre,twocolumn,groupedaddress]{revtex4-1}
\usepackage[version=3]{mhchem}
\usepackage{amsmath,amssymb,graphicx,color}
\usepackage{dcolumn}
\usepackage{bm}
\usepackage[mathlines]{lineno}
\usepackage{epsfig}
\usepackage{graphicx}
\usepackage[cp1251]{inputenc}
\usepackage[english]{babel}

\begin{document}

\title{{Enhancement of magnetic resonance imaging with metasurfaces}}

\author{\firstname{A.~P.} \surname{Slobozhanyuk$^{1,2}$},
  \firstname{A.~N.} \surname{Poddubny$^{1,3}$},
  \firstname{A.~J.~E.} \surname{Raaijmakers$^{4}$},
  \firstname{C.~A.~T.} \surname{van den Berg$^{4}$},
  \firstname{A.~V.} \surname{Kozachenko$^{1}$},
  { \firstname{I.~A.} \surname{Dubrovina$^{5}$}},
 \firstname{I.~V.} \surname{Melchakova$^{1}$},
 \firstname{Yu.~S.} \surname{Kivshar$^{1,2}$},
 \firstname{P.~A.} \surname{Belov}$^{1}$}

\affiliation{$^1$ITMO University, St. Petersburg 197101, Russia\\
$^2$Nonlinear Physics Center, Australian National University, Canberra ACT 0200,
Australia\\
$^3$Ioffe Physical-Technical Institute of the Russian Academy of Sciences,
St. Petersburg 194021, Russia\\
$^4$Department of Radiotherapy, University Medical Center Utrecht, P.O. Box 85500, 3508 GA Utrecht, The Netherlands\\
$^5${Institute of Experimental Medicine, Russian Academy of Medical Sciences, St. Petersburg 197376, Russia}}




\maketitle

\textbf{Magnetic resonance imaging (MRI) is the cornerstone technique for diagnostic medicine, biology, and neuroscience~\cite{Lauterbur1973Nature}. This imaging method is highly innovative, noninvasive and its impact continues to grow~\cite{RevModPhys1999}. It can be used for measuring changes in the brain after enhanced neural activity~\cite{Logothetis2008}, detecting early cancerous cells in tissue~\cite{fingerprinting}, as well as for imaging nanoscale biological structures~\cite{Degen2009}, and controlling  fluid dynamics~\cite{Microfluidic}, and it can be beneficial for cardiovascular imaging~\cite{heartbeatMRI}. The MRI performance is characterized by a signal-to-noise ratio, however the spatial resolution and image contrast depend strongly on the scanner design~\cite{BookMRI}. Here, we reveal how to exploit effectively the unique properties of metasurfaces for the substantial improvement of MRI efficiency. We employ a metasurface created by an array of wires placed inside the MRI scanner under an object, and demonstrate
a giant enhancement of the magnetic field by means of subwavelength near-field manipulation with the metasurface, thus strongly increasing the scanner sensitivity, signal-to-noise ratio, and image resolution. We demonstrate experimentally this effect for a commercially available MRI scanner and a biological tissue sample. Our results are corroborated by measured and simulated characteristics of the metasurface resonator, and our approach can enhance dramatically functionalities of widely available low-field MRI devices.}

It is well known that the strength of a signal delivered by conventional MRI devices depends on the static magnetic field ($B_0$) of the imaging system~\cite{BookMRI}. Over the last two decades, low-field MRI machines operating at $B_0= 1.5$~T have been used widely as key clinical tools. Recently, newly constructed high-field systems ($B_0=3$~T) are being successfully exploited in hospitals~\cite{3TeslaMRI} and there is a growing demand for ultra-high-field MRI machines (7~T or more)~\cite{9TeslaMRIResults}.

A drive for high-field magnets is fueled by the benefits of higher signal-to-noise ratio (SNR) and better image resolution~\cite{Comparison15Tand7T}. Still, to improve the overall image quality one has to overcome the increasing static and radio-frequency (RF) magnetic field inhomogeneities that lead to the susceptibility artifacts and contrast variations~\cite{FieldInhomogeneity}. Further, the most important problem at high static fields $B_0\gtrsim 7$~T is a potential harm to a  human body~\cite{SafetyMRI}. In particular, a possibility for tissue heating negligible at magnetic fields around 1.5 Tesla becomes substantial at higher fields due to an increase of the RF energy absorption~\cite{SafetyMRI}. Also, an exposure to magnetic fields above 2~T can produce effects like vertigo and nausea, therefore the examinations above {3~T} are conducted under especially careful medical supervision. Moreover, there exists quite a number of implants and medical devices that can be safe for low fields (1.5~T) and become unsafe at 3~T systems~\cite{SafetyComparison3tand15T,MRIsafetyreview}. Therefore, the problem of enhancing the image quality without increasing the static magnetic field is very relevant to the MRI technology.

\begin{figure}[b!]
\protect\includegraphics[width=1\columnwidth]{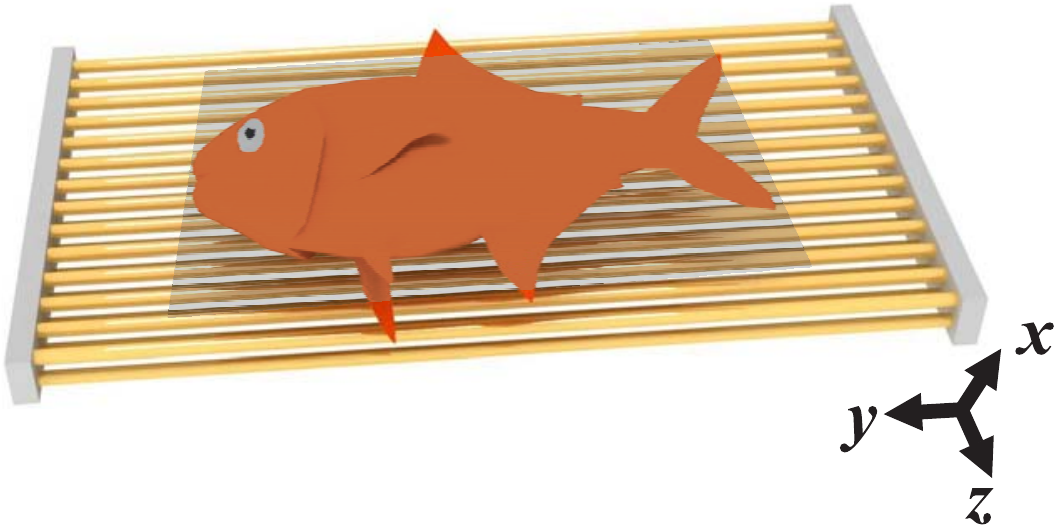}
\caption{\label{Figure1_NEW}
Artist's view of a biological object placed on a metasurface resonator. The shaded rectangular region marks the region of interest.}
\end{figure}

Up to now, many different approaches have been proposed in order to enhance the MRI characteristics, while keeping the static field unchanged. The first way is the coil optimization. The progress in the RF coil technology over the last decades has already made an essential contribution on the MR scanner design. Comprehensive work has shown that it is possible to reduce scanning time with the help
of parallel imaging methods~\cite{Sodickson,Pruessmann}, that a substantial improvement in the SNR is available through multi-channel coils~\cite{DesignApproachWithcoil}, and that larger area can be examined using traveling-wave technique~\cite{travelling-wave1,travelling-wave2}. The second approach is based on the idea to employ special contrast agents~\cite{kuperman} locally enhancing the magnetic field, such as rare-earth magnetic atoms~\cite{Thierry2005} and magnetic nanoparticles~\cite{Hogemann2000}. The third approach relies on special additional pads placed near the object. This can be done in several ways: by using high-permittivity pads in order to improve the RF field homogeneity and to reduce the specific absorption ratio~\cite{ReduceSarwithPads}; by using dielectric resonators (e.g. the cylindrical resonator when it is used as a coil with a hole in the center for an object) to increase the RF magnetic field and the SNR locally ~\cite{WebbFirstresonator,ImproveWithResonators}. However there exist several limitations for applications of the dielectric resonators in MRI technology. These limitations include large geometrical dimensions of resonators at low frequencies, difficulties with placing an object at the maximum of the magnetic field inside the resonator, and difficulty to enhance only the RF magnetic field at the surface of the resonator, which is necessary for
investigating large objects.

\begin{figure*}[t!]
\includegraphics[width=2.05\columnwidth]{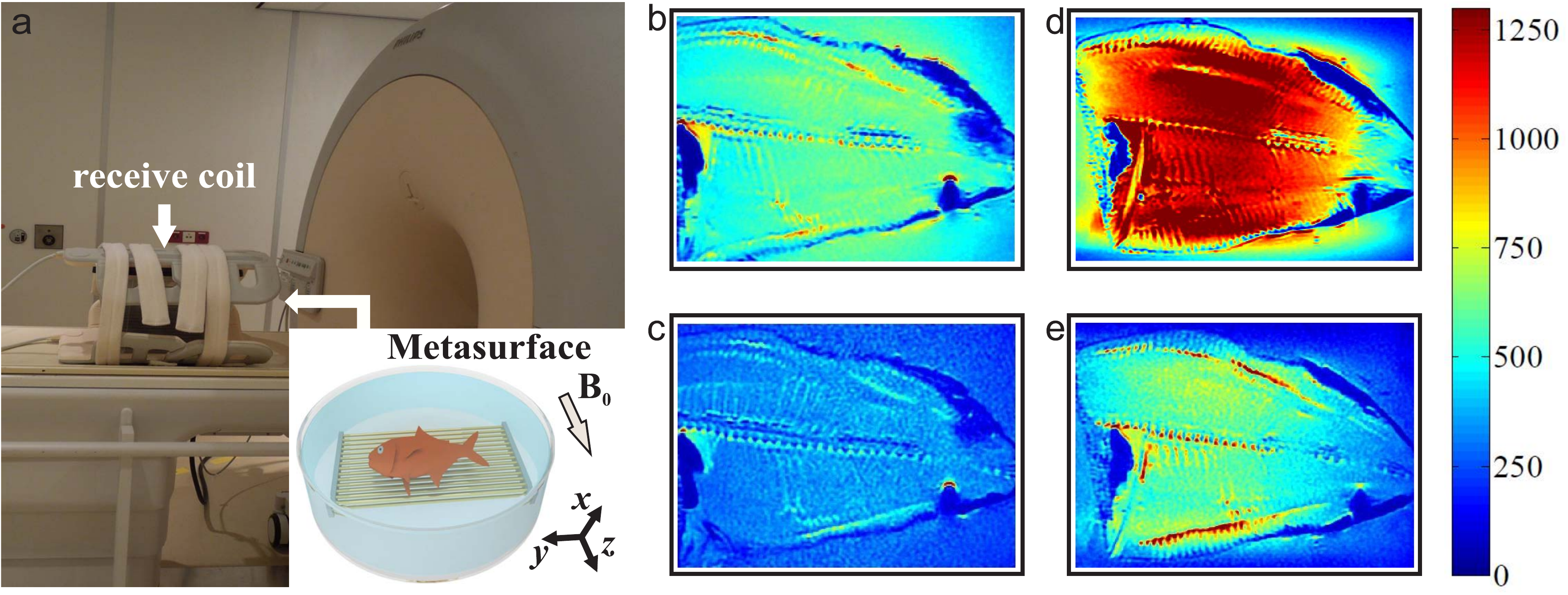}
\caption{ (a) A photo of the experimental setup placed into a MRI scanner. The inset illustrates schematically a metasurface embedded in the phantom and a sample. (b,c) MRI images (false colors) scanned without a metasurface for 1020~s and 120~s, respectively. (d,e) MRI images scanned with the metasurface for 1020~s and 120~s, respectively.
 }\label{Figure2_NEW}
\end{figure*}

The MRI efficiency can be further increased by using metamaterials, artificial structured media characterized by the values of
effective permittivity and permeability not commonly available in Nature. Metamaterials demonstrate many fascinating fundamental properties, and they are very promising for engineering the properties of electromagnetic devices and for tailoring the electromagnetic field-matter interactions~\cite{EnghetaBook06,ShalaevBook}. MRI is a very important field for applications of metamaterials. Several existing implementations of metamaterials in MRI have been already proposed, such as  Swiss-roll arrays~\cite{Wiltshire2001}, split-ring arrays with negative magnetic permeability~\cite{freire2008,Freire201081}, magnetoinductive waveguides~\cite{SymsRRA2010}, and endoscopes from arrays of metallic wires~\cite{Radu2009}. However, we believe that the full potential of the metamaterials for MRI technique
is far from being uncovered.

Here we demonstrate how 1.5~T MRI scanning efficiency is improved when the wire metamaterial-based surface (metasurface) structure is placed inside the scanner. Metasurfaces were studied by many groups~\cite{Holloway_review,ShalaevReviewScience2013} and have shown spectacular applications in the long-wavelength regime for communication~\cite{Holloway_review} and in optics  to control the propagation of light and to provide subwavelength imaging~\cite{CapassoScience2011,Alu2012NatCom}.
We propose a metasurface, based on the resonant array of metallic wires, that is inserted into the water based holder [Fig.~\ref{Figure1_NEW}]. We demonstrate that this metasurface can controllably redistribute the RF electromagnetic field at deep subwavelength level and simultaneously increases the SNR more than \emph{twice}. The immediate result of this increase is that a higher resolution image can be obtained over the same time slot or faster examination can be performed with similar resolution. Such metasurface presents a substantial step towards the improvement of MRI systems that are widely available in hospitals, without increasing the static magnetic field $B_0$.

Our main results and experimental setup are summarized in Fig.~\ref{Figure2_NEW}. We have realized a metasurface  as an array of  $14\times 2$ wires. The studied biological sample (ex vivo fish) has been placed on the metasurface structure, embedded inside the water phantom, see the inset of Fig.~\ref{Figure2_NEW}a, and placed in a 1.5 T MR scanner, that has operating frequency 63.8 MHz [Fig.~\ref{Figure2_NEW}a]. The birdcage coil, which resides in the bore of the scanner, has been used as a transmitting antenna {see Supplementary Materials and Methods for more details}. The radio frequency signal has been received by a surface body array. The scans have been performed according to the standard magnetic resonance imaging protocol (see Methods). Panels (b)-(e) summarize the obtained results. Fig.~\ref{Figure2_NEW}b and Fig.~\ref{Figure2_NEW}c present the images obtained without the metasurface structure for the long scanning time of $1020$~s (b) and for the short time of $120$~s (c). {The SNR ratio is clearly increased for longer scanning time due to the accumulation of the signal. Panels (d) and (e) present the scans obtained for the fish placed on the metasurface for the long ($1020$~s) and short ($120$~s) scanning time correspondingly. The amount of received valuable signal has substantially increased with the metamaterial [cf. Fig.~\ref{Figure2_NEW}b,c and Fig.~\ref{Figure2_NEW}d,e]. Moreover the amount of received signal for the short time with metasurface is comparable with that for longer time without the metamaterial [cf. Fig.~\ref{Figure2_NEW}b and Fig.~\ref{Figure2_NEW}e]. Hence, Fig.~\ref{Figure2_NEW} indicates, that it is possible to achieve higher signal and better image quality at lower scan time. Due to the fact that MRI image quality is evaluated by SNR, we have performed more detailed experiments and quantitative numerical simulations, discussed below.}

First, we have made experimental measurements of the SNR in the region of interest (see the black plane of Fig.~\ref{Figure1_NEW}) under the metasurface. Based on results of the separate scans of signal and noise in two configurations (phantom with the metasurface {with and w/o the RF pulse} and  empty phantom
{with and w/o the RF pulse}) we determined the ratio
 SNR$_{2} /$SNR$_{1} \approx 2.7~$, where  SNR$_{2}$ corresponds to the ratio with the metasurface inside the phantom and SNR$_{1}$ corresponds to the ratio for the empty phantom (see Supplementary Materials for more details). The increase of the SNR by more than a factor of \emph{two} effectively corresponds using an MR system with a two times higher static magnetic field. Here, it is possible to improve MRI characteristics just putting the sample on the metasurface, without using the high field scanners.

The enhancement of the signal with addition of metamaterial (Fig.~\ref{Figure2_NEW}d,e) is due to the resonant coupling to electromagnetic modes of the metasurface. The length of each wire is tuned to the Fabry-P\'erot condition for the  first eigenmode at the operating frequency of 1.5 T MRI machine, i.e. $f=c/(\sqrt{\varepsilon}L)\approx 63.8~$MHz. Here, $\varepsilon{\approx 81}$ is the dielectric constant of the background media, water in our case,
and $L=25.5$~cm is the wire length. For the first Fabry-Perot mode the highest magnetic field is localized in the middle part of the surface and the electric field is localized near the edges of the wires~\cite{Slobozhanyuk2013,Fink2012}.
Hence, the observed  effect can be explained by the RF magnetic field enhancement on the wire metasurface.
In particular, the SNR ratio is determined as a ratio of the RF magnetic field ${B_{1}^-}$ in the scanned region (a signal) to the square root of the absorbed power (a noise)~\cite{kuperman}. The Fabry-Perot resonance leads not only to the overall enhancement of both ${B_{1}^-}$ field and noise in the phantom, but it pins also the magnetic field to the scanned region suppressing the electric field and, hence, the absorbed power in this region. Such a relative redistribution of the spatial fields leads to a growth of SNR. It should be mentioned that the metasurface enhances also the transmitted RF magnetic field ${B_{1}^+}$ (see Supplementary Materials and Methods for more details).

\begin{figure}[t!]
\protect\includegraphics[width=1\columnwidth]{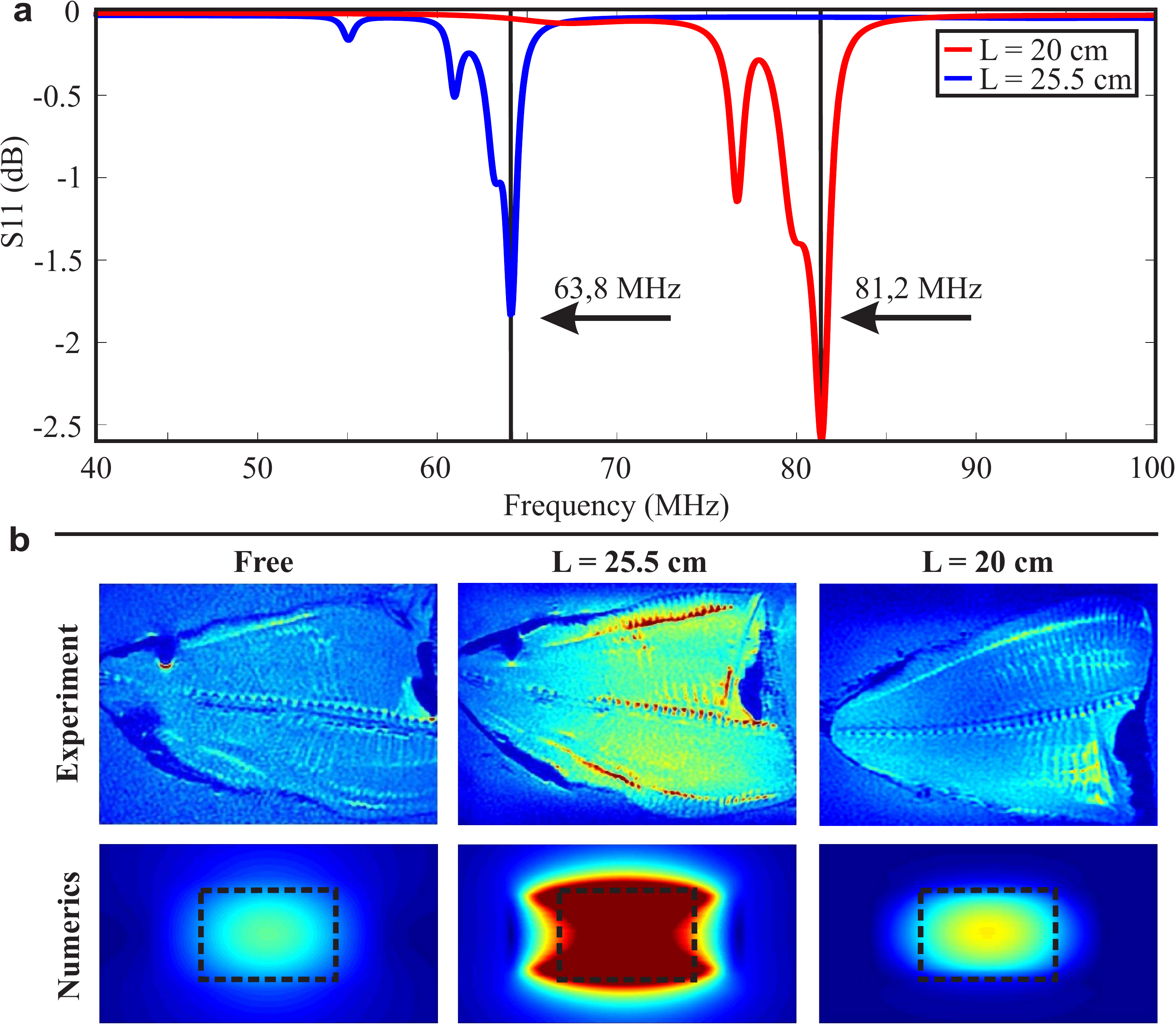}
\caption{\label{Figure4_NEW}{(a) Spectral dependence of the reflection coefficient for the magnetic loop antenna placed above the metasurface in the top plane center. Red and blue curves present the simulated spectra and correspond to the two metamaterial structures with different wire length, $L=20~$cm and $L=25.5~$cm, respectively. (b) Experimental MRI scans and numerically calculated SNR maps for three cases: without metamaterial, with metasurface length $L=25.5~$cm and $L=20~$cm respectively.}}
\end{figure}

To further confirm that the MRI enhancement of the SNR is mediated by the Fabry-Perot modes of the metasurface, we have compared {experimentally and numerically} two prototypes with different wire lengths $L=25.5$~cm and $L=20$~cm. In order to determine the working frequency of metamaterial in numerical simulation we have placed a small magnetic loop as a source under the metasurface in the center and analyzed the source reflection coefficient $S_{11}$. {The loop was located at 1.2~cm distance from the surface of the metamaterial}. The minimum of the reflection coefficient is reached at the metasurface resonance, when the power is efficiently  transmitted to the metasurface. In our case this takes place at the frequency of the first eigenmode of the metasurface. Red and blue curves in Fig.~\ref{Figure4_NEW} correspond to the reflection coefficient of the antenna for different wire lengths, $L=25.5~$cm (blue) and $L=20$~cm (red). The Fabry-Perot mode frequency and the minimum of the reflection coefficient exhibit red shift for longer wires. The Fabry-Perot resonance of the  structure with longer wires is tuned to the nuclear magnetic resonance at $f=63.8~$MHz. {This explains the observed enhancement of the MRI image, as shown in the middle column of Fig.~\ref{Figure4_NEW}b. For shorter wires, the Fabry-Perot resonance is detuned. The image quality measured for this structure is much weaker (see the right column in Fig.~\ref{Figure4_NEW}b) and very similar to the results of the scanning of the fish without a metasurface (see the left column in Fig.~\ref{Figure4_NEW}b). The subtle difference between the images without a metamaterial and with a short metasurface can be explained due to the excitation of different transverse modes of a finite metamaterial sample at the frequencies below the Fabry-Perot resonance ~\cite{SlobozhanyukAPL_2014}. The excitation of these modes leads to the redistribution of the electromagnetic field and a slight enhancement of SNR (see the right bottom panel in Fig.~\ref{Figure4_NEW}b). For even shorter wires this effect will be minimized further at the 1.5~T operational frequency and no difference between scans with short wires and without metamaterial is observed. Therefore, the detuned metasurface does not influence the imaging process. Thus, Fig.~\ref{Figure4_NEW}b conclusively demonstrates, that the experimentally observed MRI enhancement is due to the spectral matching between the first Fabry-Perot mode of the metasurface and the frequency of the nuclear magnetic resonance.}

\begin{figure*}[t!]
\includegraphics[width=2\columnwidth]{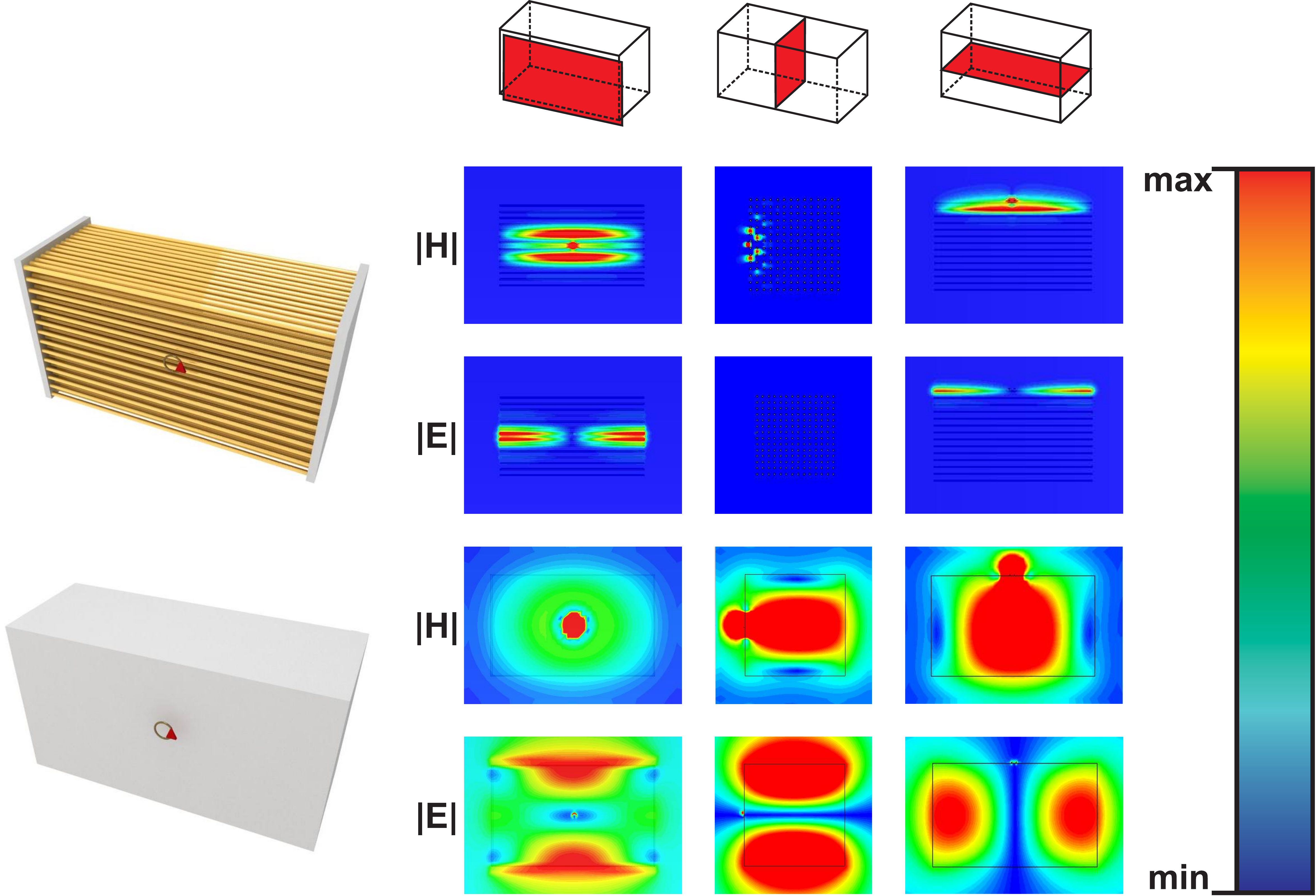}
\caption{Comparison between electric and magnetic field distributions in two resonators: wire metamaterial (on the top) and dielectric resonator (on the bottom). Left row illustrates the high-permittivity and metamaterial resonators that have been used in numerical simulations. Top line shows the schematic cut planes.
 }\label{Figure5_NEW}
\end{figure*}

The safety aspects in MRI are mainly determined by the specific absorption ratio, that quantifies the absorption of electromagnetic energy and therefore the risk of tissue heating due to the application of the RF pulses, necessary to produce the MR signal. The specific absorption ratio is proportional to the square of the induced electric field. The electric field on the surface of the metamaterial is concentrated at the edges of the wires (see the absolute value of electric field induced on the top of the metasurface Fig.~\ref{Figure5_NEW}) and has an area of minimum values in the region of interest. Therefore, as soon as the object is placed on the resonator in a proper way the examination is safe and even safer than without resonator. In our case the optimal region coincides with the region of interest (black rectangular plane in Fig.~\ref{Figure1_NEW}) that is equal to 67\% of the resonator length (see Supplementary Materials for detailed discussion). This metasurface resonator provides the possibility to homogeneously enhance and redistribute the RF field in such a way, that the magnetic field is located in the area of the object and the electric field is removed from the region of interest. Moreover, due to the low electric field region the SNR enhancement does not depend on the different tissue permittivities of various bodies. Also, the enhancement principle differs qualitatively from the earlier demonstrated metamaterial designs, mainly used to transmit an RF signal from the MRI tube or as a matching device between a coil and a patient\cite{Wiltshire2001,Radu2009,freire2008}. Here, we enhance locally only the magnetic field in the scanned region while keeping the electric field and the SAR low. Although a metasurface made of split-ring resonators\cite{freire2008} could be employed potentially in the similar regime, it suffers from stronger electric field and more inhomogeneous magnetic field in the imaged area. Hence, the proposed structure is more advantageous for a number of reasons.

With a view to underline the advantages of the metasurface resonator we have compared the proposed structure with the conventional high-permittivity resonator. The intrinsic property of the solutions to the Maxwell equations in the homogeneous medium is the coincidence between the antinodes of the magnetic field and the nodes of the electric one. Hence, it is impossible to maximize the RF magnetic field at the surface of the dielectric resonator while keeping the electric field away from the object under examination. We have numerically studied a bulk rectangular dielectric resonator with  parameters tuned to obtain the first mode at the same frequency $f=63.8$ MHz. In order to excite the resonator we have placed a magnetic loop antenna in the center near the surface (see bottom left corner of Fig.~\ref{Figure5_NEW}). For qualitative comparison we have increased the number of layers in the metamaterial, so that the bulk dielectric resonator and the metamaterial resonator have approximately the same dimensions and are excited identically (see left corner of Fig.~\ref{Figure5_NEW}). Importantly, the metasurface behaves as a bulk wire metamaterial already for two or more layers of wires~\cite{SlobozhanyukAPL_2014}.

As can be seen in Fig.~\ref{Figure5_NEW}, for a wire metamaterial resonator (top panels) the magnetic field is always localized at the resonator surface, while for a dielectric resonator (bottom panels) the magnetic field is localized inside the high permittivity structure, and electric field is localized at the surface. That is why the dielectric resonator can be employed only in the case when the object under study is placed inside of a high-permittivity resonator, which seems to be quite challenging.

We have realized an unique metamaterial resonator in the form of a wire metasurface. The Fabry-Perot resonance of the wire array is spectrally tuned to the nuclear magnetic resonance. This allows us to observe a substantial enhancement of the signal-to-noise ratio enabling a reduction of the MRI scanning time and also obtain higher image resolutions. Also it provides an opportunity to locally control the electric and magnetic fields near the object. There exists quite a number of possible implementations of metasurfaces for medical examinations inspired by the concepts of smart clothes and wearable electronics ({see Supplementary Materials for details}). For example, metal strips can be printed on clothes, so that a patient will have to wear a special lightweight jacket before the examination. In this particular case, the metasurface produces a homogeneous excitation from all sides. This involves no health risks provided that the strip ends are isolated from the patient, while it gives a possibility to use any desirable radio-frequency coils positioned near the patient. The metasurface can be embedded in certain parts of the patient table as well. To summarize, the use of metasurface is a promising pathway to improve the MRI efficiency and design simple yet effective scanning protocols.

\section*{Methods}\label{sec:methods}

All experiments were performed using a 1.5 T whole-body MRI system (Philips Healthcare) in combination with the 16 channel Torso XL receive array. The fish was located inside the water phantom and the receive array was placed around the water phantom (Fig.~\ref{Figure2_NEW}a). Two identical measurements (with the same position of the receiving coil) were performed: (1) a fish inside the water phantom without a metasurface; (2) a fish inside water phantom on a metasurface.

The metasurface was made from nonmagnetic brass wires. Wire length $L$, radius $r$ and lattice period $a$ are equal to $25.5$~cm, $0.1$~cm and $1$~cm, respectively. The wires were supported by a double-layer plastic film, fillable by water. The period of the structure was optimized to achieve a homogeneous magnetic field enhancement in the region of interest. The number of layers was chosen to reach the highest magnetic field enhancement, while keeping the thickness of the structure as thin as possible.

Initially, we have determined by the direct measurements of the $B_0$ map that there is no influence of the metasurface on the main static magnetic field of MRI. The scans shown in Fig.~\ref{Figure2_NEW} were obtained using a standard spin-echo sequence, with the following parameters: field of view was 300 mm by 300 mm, flip angle was $90\,^{\circ}$, the echo time (TE) was 20 ms, for long time scans (Fig.~\ref{Figure2_NEW}b,d) the repetition time (TR) was 1240 ms and for short scan time (Fig.~\ref{Figure2_NEW}c,e) the TR was 480 ms. For SNR measurements we obtained separate data for signal and noise. The measurement sequence was carried out twice: the first time with the RF pulses and therefore acquiring signals and noise and the second time without any RF pulses and therefore acquiring only noise. The signal was defined as the mean pixel intensity value in a region of interest, while the noise was defined as the standard deviation in this region in the noise image. For this measurement we used gradient-echo sequence (GRE), the flip angle was $15\,^{\circ}$, the TR was 50 ms and TE was 3.7 ms. After an addition of the metamaterial, the RF power levels that previously generated $90\,^{\circ}$ excitation pulses in the region of interest created a much larger angle due to enhancement of the RF field. In order to avoid over-tipping with metasurface present, the power level of the excitation pulses were re-optimized (reduced).

Numerical simulations were performed using the time-domain solver of CST Microwave Studio 2013 package. The small brass ring, excited by the discrete port (see red triangle in the left schematic plots of Fig.~\ref{Figure5_NEW}) was used as a source.

\section*{Acknowledgements}\label{sec:Acknowledgements}

The authors are grateful to M. Kozlov, A.V. Shchelokova, M. Lapine, M. Barth, J.D. Baena, A.E. Miroshnichenko, I.V. Shadrivov, D.A. Powell, C.R. Simovski, and S.A. Tretyakov for useful discussions and suggestions. This work was supported by Government of Russian Federation (Grant 074-U01), Dynasty Foundation (Russia), and the Australian Research Council. ANP has been supported by the European Commission project ``SPANGL4Q''.

\section*{Author Contributions}\label{sec:Author Contributions}
Numerical analysis was carried out by A.P.S and experiments were carried out by A.P.S., A.J.E.R., C.A.T.v.d.B., {I.A.D} and A.V.K. All authors analyzed and discussed the results. The manuscript and figures were prepared by A.P.S. and A.N.P.

\bibliography{literature}

\end{document}